\documentclass[conference]{IEEEtran}
\usepackage{cite}
\usepackage[cmex10]{amsmath}
\usepackage[tight,footnotesize]{subfigure}
\usepackage[font=footnotesize]{subfig}
\usepackage{url}
\usepackage{lipsum}
\usepackage{multicol}
\usepackage{graphicx}

\author{\IEEEauthorblockN{
Andre Recnik\IEEEauthorrefmark{1},
Kevin Bandura\IEEEauthorrefmark{3}, 
Nolan Denman\IEEEauthorrefmark{1}\IEEEauthorrefmark{2},
Adam D. Hincks\IEEEauthorrefmark{4}, 
Gary Hinshaw\IEEEauthorrefmark{4} \\
Peter Klages\IEEEauthorrefmark{1},
Ue-Li Pen\IEEEauthorrefmark{5},
and Keith Vanderlinde\IEEEauthorrefmark{1}\IEEEauthorrefmark{2}}
\IEEEauthorblockA{\IEEEauthorrefmark{1}Dunlap Institute for Astronomy \& Astrophysics, University of Toronto}
\IEEEauthorblockA{\IEEEauthorrefmark{2}Department of Astronomy \& Astrophysics, University of Toronto}
\IEEEauthorblockA{\IEEEauthorrefmark{3}Department of Physics, McGill University}
\IEEEauthorblockA{\IEEEauthorrefmark{4}Department of Physics and Astronomy, University of British Columbia}
\IEEEauthorblockA{\IEEEauthorrefmark{5}Canadian Institute for Theoretical Astrophysics, University of Toronto}
Contact Email: vanderlinde@dunlap.utoronto.ca
}

\title{An Efficient Real-time Data Pipeline for the CHIME Pathfinder Radio Telescope X-Engine}

\begin{document}
\bstctlcite{kernelBSTcontrol}

\maketitle

\begin{abstract}
The CHIME Pathfinder is a new interferometric radio telescope that uses a hybrid FPGA/GPU FX correlator. The GPU-based X-engine of this correlator processes over 819\,Gb/s of 4+4-bit complex astronomical data from N=256 inputs across a 400\,MHz radio band.  A software framework is presented to manage this real-time data flow, which allows each of 16 processing servers to handle 51.2\,Gb/s of astronomical data, plus 8\,Gb/s of ancillary data.  Each server receives data in the form of UDP packets from an FPGA F-engine over the eight 10\,GbE links, combines data from these packets into large (32MB-256MB) buffered frames, and transfers them to multiple GPU co-processors for correlation.  The results from the GPUs are combined and normalized, then transmitted to a collection server, where they are merged into a single file.  Aggressive optimizations enable each server to handle this high rate of data; allowing the efficient correlation of 25\,MHz of radio bandwidth per server. The solution scales well to larger values of N by adding additional servers.
\end{abstract}

\section{Introduction}

The increasing performance of commodity computer hardware has opened up new possibilities for building real-time radio correlators.  Traditionally correlators have used custom ASICs or FPGAs for all calculations. The first systems built with off-the-shelf hardware used CPUs, for example the real-time software correlator designed for the Giant Metrewave Radio Telescope (GMRT) \cite{gmrt}.  While early experiments using GPUs showed relatively poor performance \cite{manycore}, the development of new GPUs and the efficient xGPU code \cite{xgpu} for NVIDIA's CUDA platform has popularized the use of GPUs in FX style correlators.

FX correlators operate in two stages: first the F-engine samples astronomical data from each radio input and channelizes it into frequency bands using a Fourier transform, then the X-engine correlates all of the inputs against one another within each frequency band.  The low cost and flexibility of 10 Gigabit Ethernet (10\,GbE) has enabled these two stages to be easily separated, as demonstrated by Parsons, et al. \cite{packet-switched}; with separate FPGAs performing the F-engine and X-engine stages connected by a 10\,GbE packet switched network.  A recent trend has been to replace FPGAs with GPUs in the X-engine, leading to the so called hybrid correlator approach.  Projects using  hybrid correlators include: the Precision Array for Probing the Epoch of Reionization (PAPER) \cite{paper}, the Large Aperture Experiment to Detect the Dark Ages (LEDA) \cite{leda}, and the Murchison Wide-field Array (MWA) \cite{mwa}.


The Canadian Hydrogen Intensity Mapping Experiment (CHIME) Pathfinder \cite{bandura14} is a cylindrical radio telescope with 128 dual-polarization receivers for $N=256$ total inputs.  It uses a newly developed hybrid FX correlator.  The F-engine consists of 16 custom FPGA boards, each with 16 ADCs, which sample and channelize the data into 1024 frequency bins.  The data are reduced to 4+4-bit complex numbers, then a custom backplane network shuffles this data, such that each FPGA has data for all inputs in a subset of frequency bands.  The shuffled data is then transmitted over 128 x 10\,GbE links in User Datagram Protocol (UDP) packets to a GPU based X-engine.  

This paper presents the software pipeline, called \texttt{kotekan},\footnote{Given the musical acronym of the experiment, CHIME, the collaboration uses musically inspired names for system components.  Kotekan is a style of playing fast interlocking parts in Balinese Gamelan music using percussive instruments. \url{http://en.wikipedia.org/wiki/Kotekan}} which manages data flow in the X-engine. The data is received from the F-engine as UDP packets, is merged and sent to the GPUs for processing, then the output from the GPUs is sent out to the collection and aggregation server.
The software is written in the C programming language.

\begin{figure}
  \includegraphics[width=\columnwidth]{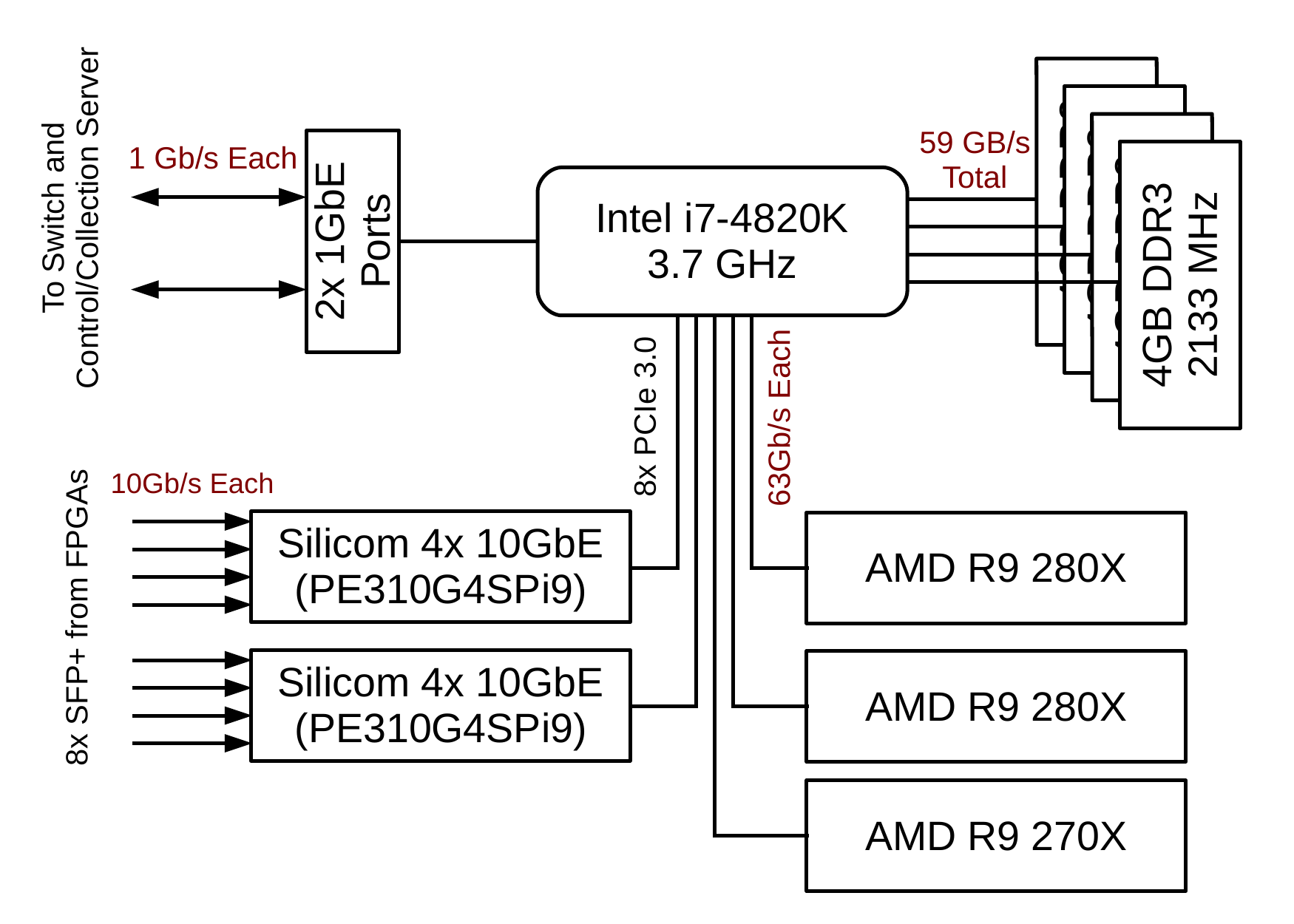}
  \caption{Abstract GPU server layout.  Speeds are the theoretical transfer rates supported by the given bus; achievable rates vary.  There are 16 GPU servers in the in the CHIME Pathfinder.}
  \label{fig:abs_layout}
\end{figure}

In the CHIME Pathfinder, this software runs on 16 servers, built mostly with low-cost consumer-grade hardware.  Each server has one Intel i7 4-core CPU, two AMD R9 280X GPUs, one AMD R9 270X GPU, two Silicom 4x 10\ GbE network interface cards (NICs), and 16\ GB of DDR3 RAM.  The GPUs and NICs are each connected to the CPU by 8x PCI Express (PCIe) 3.0 lanes. An abstract layout of each of these components is given in Figure~\ref{fig:abs_layout}.  The operating system used is CentOS~Linux~6.5.

\section{Data Flow}

\begin{figure}
  \includegraphics[width=\columnwidth]{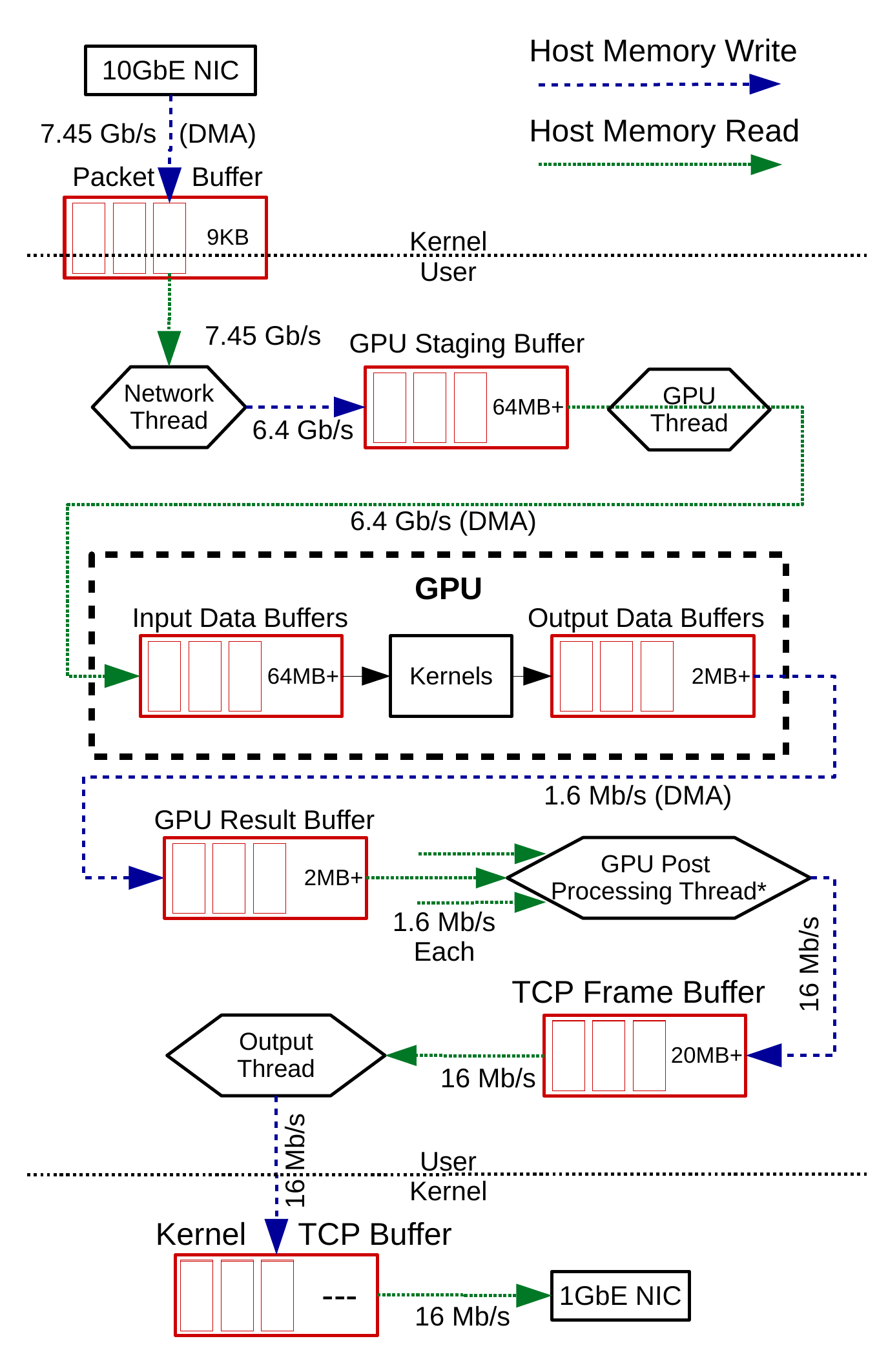}
  \caption{Data flow for one of eight streams per server.  This chart traces data from one of the 10GbE inputs.  GPU input data rates are constant, while output data rates depend on the desired integration time and are shown here for 10 seconds.  The values beyond the ``GPU Post-Processing Thread'' represent a merged stream.} 
  \label{fig:data_flow}
\end{figure}

The software design is based around generic buffer objects, which support multiple consumers and producers, and handle the majority of thread synchronization tasks using Linux \texttt{pthreads}.  In \texttt{kotekan}, they are almost exclusively used as FIFO ring buffers.  Most subsystems in \texttt{kotekan} communicate only via these ring buffers, allowing individual components to be changed without affecting the system as a whole.  This also allows individual components to have variable run times, without negatively impacting other operations.

The main components in \texttt{kotekan} are networking threads, GPU threads, GPU call-back threads, a GPU post-processing thread, and an output thread.  Each of these components interfaces with one or more of the buffer objects.   These threads form the data path as shown in  Figure~\ref{fig:data_flow}.

Since the data changes shape between steps, an ancillary data package is carried along each step of the data path.  To limit calls to \texttt{malloc} and \texttt{free}, these objects live in a pool which is only allocated at the start of the application, and only freed at the end of the application. In fact, the entire application does not \texttt{malloc} or \texttt{free} any memory outside of startup and shutdown.

Each of the eight network inputs is processed as its own stream until after the correlation, at which point it is combined with the other 7 streams on the system before being sent to the collection server.   In the current configuration, each of the AMD R9 280X GPUs processes 3 of these streams, and the AMD R9 270X processes 2.  The mapping of streams to GPUs can be adjusted by configuration file.

In addition to \texttt{kotekan}, which runs on the processing servers, software on the data collection server combines results from all of the processing servers into a single file.   

The following subsections trace this data flow from the F-engine through to the output file.

\subsection{UDP Packet Processing}

\begin{figure}
  \includegraphics[width=\columnwidth]{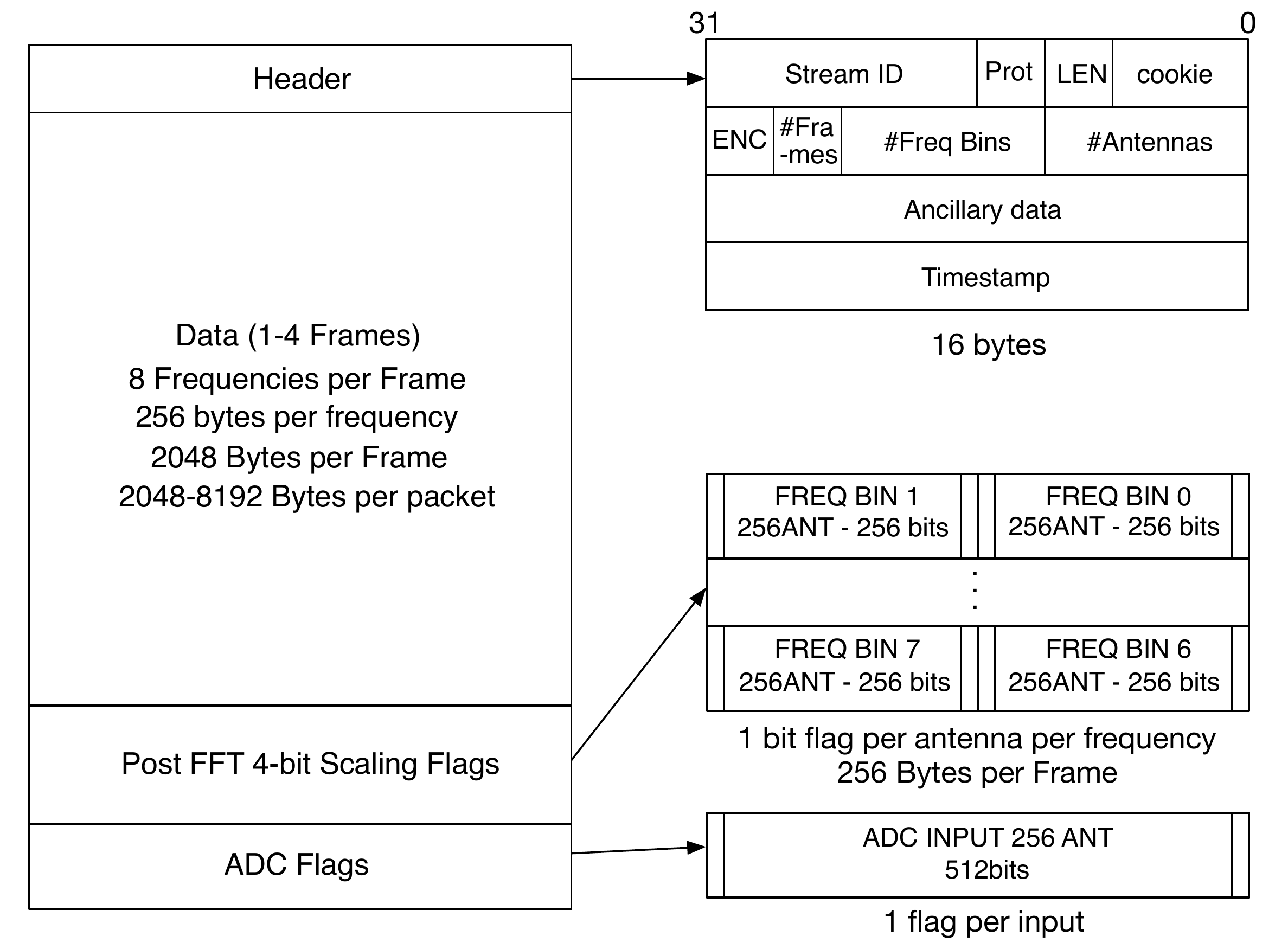}
  \caption{The UDP packet format sent by the FPGAs.} 
  \label{fig:packet_format}
\end{figure}

Channelized data is received from the F-engine in UDP packets over standard 10Gb/s Ethernet lines (10GbE).  Each UDP packet is $\sim$9K in size and has the structure shown in Figure~\ref{fig:packet_format}.  The packet is designed in such a way that the channelized data section can be used by the GPU without modification.  All that is required of the system is to place that data into the right location in a large host frame, which can then be copied to the GPU over PCIe.  While the CPU does have to process the header and footer information, it very rarely needs to modify the data. 

With $N=256$ inputs, the F-engine packs all inputs for 8 frequency bands into each 10\,GbE stream.  Each of the values is a 4+4-bit complex number, so all 256 inputs for one frequency take up 256\,bytes.  For 8 frequency bands this represents 2KB.  To reduce the number of packets per second and lower the CPU load, 4 time samples are recorded in each packet.  The header contains a \texttt{streamID} that identifies which frequency bands are represented in the packet, and a sequence number which gives the relative time and is synchronized across the entire array.

There are a number of considerations to make when processing UDP packets at high rates:

\subsubsection{Efficiently receiving packets from the network}

The Linux kernel has a large overhead when processing high bandwidth network traffic.  Early tests showed that even with large packets and optimizations to the Linux kernel network parameters, it would be unable to process the required data rates while simultaneously managing the GPU PCIe transfers.  The solution was to bypass the Linux kernel's network stack entirely using modified network drivers and custom kernel modules.  There are a number of pre-built solutions which enable kernel bypass, including: DPDK \cite{dpdk}, netmap \cite{netmap}, and others.  For this project, NTOP's PF\_RING/DNA \cite{pfring, dna} framework was chosen for its support on the chosen network card vendor.  However, the overall system is not tied to a particular bypass stack; changing the bypass stack would just involving replacing calls to the PF\_RING API with another bypass API.

The most important aspect of these bypass drivers, aside from avoiding slow kernel packet processing code, is the use of a co-mapped ring buffer that is addressable in both user and kernel space.  This allows the NIC driver to write directly to the co-mapped memory with a Direct Memory Access (DMA) operation, avoiding the traditional kernel-to-user-space copy.

\subsubsection{Handling Losses/Errors}

Since the network protocol is UDP, the system must handle lost, out of order, and duplicate packets.  This is achieved by tracking the sequence number in each packet header.  When the system detects packet loss, it writes zeros into the area of the GPU staging buffer where the missing packet would have been placed.  This removes the requirement to zero the buffers between each use; the network thread guaranties the buffer will have good data or zeros in every location.  In the case of out of order packets, the sequence number is used to copy the packet into the correct location.  Duplicate packets are simply ignored.

In early testing with the standard socket API, vectored I/O was used to separate the packet into data and header components with a \texttt{readv} system call.  The sequence number was inferred based on the pervious packet, and the data was copied directly to the GPU staging buffer.  If the sequence number did not match the expected number, then the data was moved or overwritten.  Given low packet loss, this saves a second copy for the vast majority of packets, and allowed maximum capture rates around 40\,Gb/s per CPU.  With the addition of co-mapped memory this became unnecessary, since the sequence number can be read before moving the data to the GPU staging buffer.

\subsubsection{Efficient Memory Transfers}

The standard memory copying functions like \texttt{memcpy} found in \texttt{libc} are inefficient for use in a high bandwidth environment like this one.  This issue  largely stems from the fact that compilers assume temporal locality, that data being copied will be used in the near future and should be added to the CPU cache.  This causes the destination memory to be cached in the process of performing a copy.  This puts huge pressure on the cache when moving large amounts of data, resulting in unnecessary memory controller usage.

To mediate the issue caused by the standard \texttt{memcpy}, a set of custom memory copy functions that use Intel AVX intrinsics with non-temporal hints were created.  The non-temporal hint prevents the memory copy destination from being added to the CPU's cache.  These functions copy the data from the co-mapped packet buffer to the GPU staging frame. 

\subsection{Kernel Invocation Process}

\begin{figure*}
  \includegraphics[width=\textwidth]{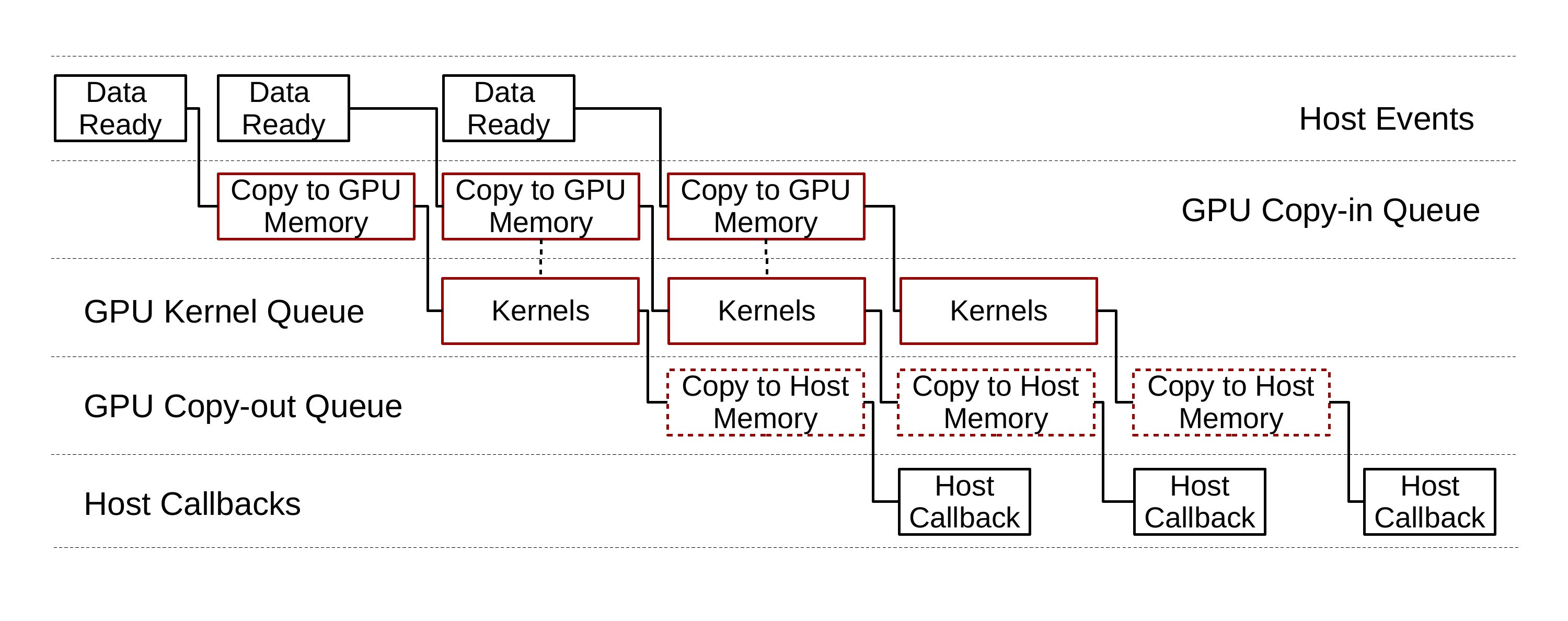}
  \caption{GPU Pipeline.  The ``Kernels'' stage includes correlation, as well as extra data processing operations like time shifting or RFI detection.  The dashed lines indicate optional steps or dependencies.  The time taken by each step is not to scale.} 
  \label{fig:gpu_pipeline}
\end{figure*}

A set of GPU programs called kernels do the actual cross-correlation and integration.  The kernels are written in the Open Computing Language (OpenCL) framework\footnote{\url{https://www.khronos.org/opencl/}}, allowing them to be run on a number of different platforms.  The kernels used achieve very high levels of efficiency using packed multiply\-accumulate operations and are detailed in \cite{peter}.

The first step in this process is copying the data to the GPU's device memory, a process largely managed by the OpenCL drivers.  In our tests we found that large (32\,MB or greater) frames transferred most efficiently.  The memory used for these frames is page-locked allowing the GPU drivers to preform an efficient DMA operation to copy the memory between the host and device. 

In contrast to some systems using xGPU \cite{xgpu} like the MWA \cite{mwa} and LEDA correlators \cite{leda} which promote the data from 4-bit to 8-bit integers in their CPUs, this system does not promote the data before sending it to the GPUs.  Avoiding this promotion reduced both CPU memory and PCIe bandwidth, and was a key part in limiting the number of servers and CPUs needed.

The entire process is pipelined to allow concurrent operation of the memory transfers and kernel invocations, as shown in Figure~\ref{fig:gpu_pipeline}.  The ``Copy to Host Memory'' step is not run on every invocation, since the correlation kernels simply add to the output of the last run, extending the integration time of the output.

When executing an OpenCL device operation, for example a memory copy or kernel invocation, there is a delay between the function call and the time the device starts the operation.  To minimize this delay, we pre-load entire chains of OpenCL operations, the first of which is set to depend on the ``Data Ready'' event in Figure \ref{fig:gpu_pipeline}, which is triggered by the GPU thread when a buffer has been filled by a network thread.  When a chain finishes, a call-back function (``Host Callback'' in Figure \ref{fig:gpu_pipeline}) adds a new chain of events which replaces the one that just finished, and marks the associated output buffer as full, so the GPU post-processing thread can take the data.

\subsection{GPU Post-Processing}

This GPU post-processing thread combines all eight streams, and then normalizes the results to correct for lost input samples.

Loss of input samples can result from packet loss, or numerical range and sampling limitations in the FPGAs and ADCs.  When this happens the data points are zeroed, before they are transferred to the GPU. This causes some correlated data points to contain fewer input samples in each integration.  The network threads track packet loss and F-engine error flags, which are used by the GPU post-processing thread to generate a normalization matrix as follows:

\begin{itemize}

  \item Packet loss is tracked as a single counter, then added to every point in the normalization matrix at the end of each integration.
  \item The F-engine flags are extracted from bit-fields in the packet footers and stored in counters for each input and frequency per integration.  The counters are used to populate the normalization matrix at the end of each integration.
  \item If two or more inputs are flaged by the F-engine in the same time sample, the previous step will result in a double counting in the normalization matrix at their intersections.  To correct this, these intersections are recorded per sample, and subtracted from the normalization matrix.

\end{itemize}

The fraction of lost data given by the normalization matrix is then applied to the correlation matrix.

This process is optimized in two ways. First the bit-field of flags in the footer is read initially as 32-bit integers, and bit-field extraction is performed only if the 32-bit representation of the bit-field is non-zero.  With a low number of flaged data points, this avoids checking each bit individually.  Second, by only updating counts of errors and their associated intersections, the number of memory accesses is greatly reduced.  The complexity of tracking the intersections per time sample is $O(E^2)$, where $E$ is the number of flaged data points in a given time sample.  This can be processed by the CPU, provided $E \ll N$.

After the data has been combined and normalized, it is formatted for transmission over a TCP stream and placed in an output buffer.  A final output thread then transmits this data to the collection server using a TCP socket connection. 

\subsection{Data Collection}

At the collection server, the TCP streams from each \texttt{kotekan} instance are combined.  The \texttt{streamID} is used to identify the frequency bands provided by each stream, allowing the collection server to correctly order the frequency bands regardless of how the 10\,GbE links were connected to the individual processing servers. The sequence number is used to align the frames in time.  The data is saved to disk using the Hierarchical Data Format Version 5 (HDF5) standard\footnote{\url{http://www.hdfgroup.org/HDF5/}}.

Individual servers may be switched off and back on without interrupting the entire system.  If one or more of the servers stops working, the collection server simply stops recording the frequency bands associated with that server.  When the server is repaired, it reconnects to the collection system, and those frequency bands resume recording.

\section{Discussion}

\begin{table}[!t]
\renewcommand{\arraystretch}{1.3}
\caption{Data Processing Rates}
\label{table_example}
\centering

\begin{tabular}{|p{2cm}|c|c|c|}
\hline
Data Type & Rate per Link & Rate per Host & Full System Rate\\
\hline
4-bit Sky Data & 6.4 Gb/s & 51.2 Gb/s & 819.2 Gb/s\\
\hline
4-bit Sky Data + Headers \& Flags & 7.45 Gb/s & 59.6 Gb/s & 953.6 Gb/s \\
\hline
Packets per Second (PPS) & 97,565 PPS & 781,250 PPS & 12.5 MPPS\\
\hline
Output Data (10s Cadence) & 16.8 Mb/s & 16.8 Mb/s & 269.5 Mb/s\\
\hline

\end{tabular}
\end{table}

This software pipeline design was largely focused on maximizing input data bandwidth per CPU, in order to minimize the overall cost of the system.  Table~\ref{table_example} shows the data rates achieved by the software in operation.

There are ways in which the pipeline could be made even more efficient.  The packets are written to a user/kernel space mapping buffer, then copied to a GPU staging buffer.  This copy could be avoided if the data segment of each packet was sent directly to the GPU buffers via DMA from the NIC, with the headers/footers sent to another buffer for processing.  Such a solution would require a tighter coupling with software and the NIC driver, so that the userspace application could direct which memory the NIC DMA transfered packet sections into.  The OpenFabrics Enterprise Distribution\footnote{\url{https://www.openfabrics.org/}} (OFED) kernel bypass software can in principle provide this; it was not pursued it due to lack of driver support on the chosen hardware.

As the number of inputs $N$ increases, the computation cost of correlation increases as $O(N^2)$, while data bandwidth scales only as $O(N)$.  With more servers required to do the correlation, the bandwidth to each should go down.  However, as co-processors become more powerful, it will be possible to correlate larger values of $N$ with fewer servers, continuing to put pressure on the bandwidth requirements of each server.

The generic design of this software allows it to be extended easily to large values of $N$.  It has already been scaled from $N=16$ in early tests of the CHIME Pathfinder using one server, to the current $N=256$ mode using 16 servers, and is expected to scale to $N=2048$ for the full CHIME experiment.

\section{Conclusion}

We have developed an optimized software pipeline to manage the data flow in the X-engine of a hybrid FX FPGA/GPU correlator.  Using a combination of kernel bypass network drivers, efficient memory copy functions, memory pools, minimal memory copies, large packets and GPU frames, and GPU kernels that can process 4-bit data, the system achieves a very high input data bandwidth per server and per CPU.   This allows input data processing rates of over 51.2\,Gb/s per single socket server, and 819\,Gb/s system wide.  In terms of radio bandwidth, the systems process the full 400\,MHz band of the CHIME Pathfinder, with $N=256$ inputs, using only 16 servers.  The system has been shown to scale well from $N=16$ to $N=256$, and will be used in a future system with $N=2048$ and a total input bandwidth of close to 8\,Tb/s.

\section*{Acknowledgments}

We are very grateful for the warm reception and skillful help we have received from the staff of the Dominion Radio Astrophysical Observatory, which is operated by the National Research Council of Canada.
We acknowledge support from the Canada Foundation for Innovation, 
the Natural Sciences and Engineering Research Council of Canada, 
the B.C. Knowledge Development Fund,  
le Cofinancement gouvernement du Qu\'ebec-FCI, 
the Ontario Research Fund, 
the CIfAR Cosmology and Gravity program, 
the Canada Research Chairs program, and 
the National Research Council of Canada.
We thank Xilinx University Programs for their generous support of the CHIME project,
and AMD for donation of test units.  Peter Klages thanks IBM Canada for funding his research and work through the Southern Ontario Smart Computing Innovation Platform (SOSCIP).

\bibliographystyle{IEEEtran}
\bibliography{cite}

\end{document}